# Integrating citizen science with online learning to ask better questions

**Vineet Pandey, Scott Klemmer**

Design Lab, UC San Diego
{vipandey, srk}@eng.ucsd.edu

**Amnon Amir, Justine Debelius, Embriette R. Hyde, Tomasz Kosciolek, Rob Knight**

Department of Pediatrics, UC San Diego
{amamir, jdebelius, ehyde, tkosciolek, robknight}@ucsd.edu

**Abstract**

Online learners spend millions of hours per year testing their new skills on assignments with known answers. This paper explores whether framing research questions as assignments with unknown answers helps learners generate novel, useful, and difficult-to-find knowledge while increasing their motivation by contributing to a larger goal. Collaborating with the American Gut Project, the world's largest crowdfunded citizen science project, we deploy *Gut Instinct* to allow novices to generate hypotheses about the constitution of the human gut microbiome. The tool enables online learners to explore learning material about the microbiome and create their own theories around causal variances for microbiome. Building on crowdsourcing or serious games that use people as replaceable units, this work-in-progress lays our plans for how people (a) use their personal knowledge (b) towards solving a larger real-world goal (c) that can provide potential benefits to them. We hope to demonstrate that *Gut Instinct* citizen scientists generate useful hypotheses, perform better on learning tasks than traditional MOOC learners, and are better engaged with the learning material.

## Can online learners perform useful work in citizen science projects?

Crowdsourcing scales well and provides good results when people's untrained intuitions are on average good, e.g. in tasks labeling images (von Ahn et al. 2004) and performing real-time captioning (Lasecki et al. 2012). This holds for tasks most people are *naturally expert* at, such as recognizing objects in images, or transcribing what's spoken in their language (Surowiecki 2005). However, for many tasks, people might have lousy estimates or guesses, if any. Such tasks require *domain-specific expertise* in breadth of knowledge (such as identifying a cat's breed in an image) or in understanding deeper features (such as describing the quality of a painting). In such cases, crowdsourcing tries to do useful work by training novices but the results are mixed.

### Citizen Science projects, though important, appeal to a limited set of hobbyists

*Citizen science* seeks to solve large scientific challenges using a distributed set of people to perform tasks (Bonney et al. 2009). Biology problems dominate popular online citizen science efforts, such as Foldit (https://fold.it) for protein folding, EteRNA (www.eternagame.org/) for RNA design, and Phylo (phylo.cs.mcgill.ca/) for small-scale multiple sequence alignment problems. Moreover, scientific datasets created from massive efforts like the Human Genome/Microbiome Projects are difficult to analyze due to (a) vast set of features and (b) gaps in our understanding of these topics. This interest in finding alternate ways to analyze data works well with people's native expertise in tasks such as identifying high-level patterns, used in games like Phylo. Designing learning modules for citizen science has demonstrated improved domain knowledge among participants (Lee et al. 2016).

However, most citizen science projects still provide minimal training and utilize participation towards low-cognition tasks like identifying certain objects in images. Since these topics from niche area, they interest hobbyists and do not scale to people beyond a small community. Galaxy Zoo (www.galaxyzoo.org) is such an example where space enthusiasts help classify galaxies. Recent citizen science projects, such as American Gut Project (http://americangut.org/) have pulled people in the loop as contributors: subjects who provide their own physical and behavioral data. We consider the next step of this evolution. How can we transform excited contributors into active collaborators who can generate hypotheses as well? Our key insight is that motivated contributors to a citizen science project can develop expertise using online learning material and collaboratively create novel knowledge.

### Online learning is underexplored as a platform to bring together crowds to do useful work

Online learners spend considerable time learning new skills and testing them on assignments with <u>known</u> answers. Could we better support their learning by asking them to apply their skills and fresh perspective towards citizen science problems with <u>unknown</u> answers? We test our idea in the context of the human gut microbiome research. The human gut microbiome is the community of microbes (and their gene products) interacting in the human gut. The American Gut Project (AGP) gives people the ability to contribute to microbiome science by providing samples for bacterial marker gene sequencing and analysis. Participants receive a summary of their results along with all of their raw data. The project's goal is to build a comprehensive map of the human microbiome, and identify good and bad areas on that map. Training AGP participants about the gut microbiome and having them identify associations between the microbiome and health and disease states can potentially accelerate this process. At the massive scale of MOOCs, this work can identify theories around whether people with similar habits actually demonstrate similarity in their gut microbiome as well.

## Gut Instinct: Basic System Design

To encourage people to brainstorm hypotheses about the gut microbiome, we've created an online collaborative brainstorming tool called *Gut Instinct*. Gut Instinct combines online learning material about the gut microbiome (divided into topics), rapid feedback to answer misconceptions (using expert insights), and an open board for learners to add/edit/discuss hypotheses (for collaborative work). Pilot studies demonstrated that framing hypothesis with a clear intent while drawing significant insights from others was a challenging task. Hence, we design our hypothesis in a three-level question format, as shown in Figure 1. All questions are provided tags by users that lead to specific topics with pre-curated content (e.g., "food", "eat", "pasta", "noodles" tags all redirect to "diet" page). This matching is done manually right now but it can be automated using topic modeling.

## Identifying patterns for online learning and crowd-work to assist each other

We hypothesize that doing useful work on real-world problems helps learning, and vice versa. We break down our broad *WorkLearn hypothesis* to three specific hypotheses:

1. **Learning improves quality of work on relevant problems:** For domain-specific tasks, crowd workers need training to create meaningful work, as provided by a quick tutorial or expert examples. However, for creative brainstorming in a scientific domain, which learning material should be used and how? We provide different tutorials, articles, and expert examples to see the benefits towards users' hypotheses generation task.

2. **Working on relevant real-world problems improves learning:** Problem sets and assignments get students to apply concepts in a specific context. Similar to Problem-based Learning approaches, we want to operationalize the insight that reflecting on concepts and using them makes learners aware of their strengths and limitations. How can we recreate similar set-up with real-world challenges that might also motivate people by having them contribute to a bigger goal?

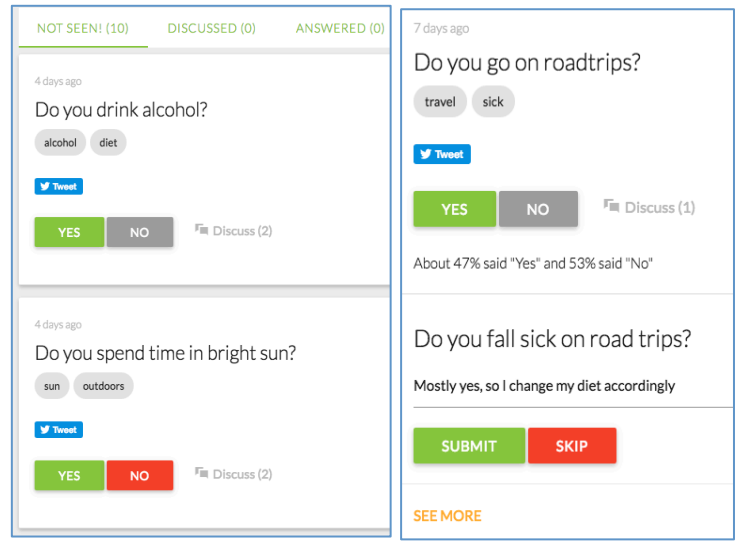

*Figure 1: (left) Gut Instinct Gutboard, where citizen scientists add and discuss questions (right) Structure of a question in Gutboard. Level 1 question is a basic yes/no question that filters out people who might not be target audience for its topic. Level 2 question invites more specific details, while the final level ("see more") invites open discussion.*

3. **Working while learning improves learners' engagement with the learning material:** Can working on real-world problems like brainstorming about causal relationships for gut microbiome engage people better than standard video lectures and forums?

Our goal is to operationalize these reasonable hypotheses for online classes and citizen science work. We are currently deploying our study to test our ideas.